\documentstyle[preprint,aps]{revtex}

\begin{document}
\preprint{KAIST-CHEP-95/16}

\title{\bf Goldstone  Supermultiplet as the Messenger
of Supersymmetry Breaking}
\author{Kiwoon Choi}
\address{
Department of Physics,
Korea Advanced Institute of
Science and Technology,
Taejon 305-701, Korea}
\maketitle

\def\be{\begin{equation}}
\def\ee{\end{equation}}

\begin{abstract}
We consider supersymmetric models in which a (pseudo) Goldstone
supermultiplet plays the role of the messenger of supersymmetry breaking.
Such models lead to a highly predictive
form of flavor and CP conserving  soft terms,
particularly the soft scalar masses and
trilinear couplings vanish
at the scale where the (approximate) global symmetry
is spontaneously broken.
We discuss also the possibility for realizing this scheme
in supergravity models derived from
string theories.
\end{abstract}


\narrowtext

Despite of its great phenomenological success
the minimal standard model of particle physics is not considered as
a final story, but as a low energy limit of some extended version.
One of the most promising extension of the standard model is to introduce
supersymmetry (SUSY) at  the weak scale \cite{nilles}.
As is well known, any realistic supersymmetric model
at the weak scale contains explicit but soft
SUSY breaking terms which are presumed to
originate from a spontaneous  SUSY breaking
at some higher energy scale.
Most of the SUSY phenomenology of the model
crucially depends upon these soft terms.
If one writes down the most general soft terms consistent with the
low energy symmetries, it would require
too many arbitrary
parameters to describe the model.
Furthermore for a generic form of soft terms,
the typical mass of superpartners
must exceed few to  ten TeV's in order to satisfy
the strong  constraints from the flavor changing
neutral current phenomena \cite{masiero}
and the electric dipole moment of the neutron \cite{edm}.
Such a large superpartner mass spoils the natural
emergence of the weak scale, and thus the major motivation
for introducing SUSY.

In view of this difficulty,
it is desirable
to have a natural scheme leading
to a predictive form of soft parameters which  conserves
both the flavor
and CP invariances. Then soft parameters
can have a size of order 100 GeV without any difficulty,
allowing the experimental  test of the predictions of
the model in the near future.
In regard to the flavor conservation, many interesting schemes
have been proposed so far
\cite{noscale,dine,dilaton,nir,ross,dimo}.
Although some of  these schemes \cite{noscale,dine,dilaton}
can shed a light on
the CP conservation also,
it is still difficult to
eliminate the dangerous CP-violating phase ${\rm arg}(B/m_a)$
without a fine tuning \cite{choi}.
(Here $B$ denotes the  Higgs mixing parameter in the scalar
potential and $m_a$'s are the gaugino masses.)
Also some of these schemes \cite{dilaton,dimo}
are based on the dynamical properties of
the model which are simply postulated.

It would be interesting that
a predictive form of flavor and CP conserving  soft terms arises
as a consequence of some symmetries
of the model in a way  {\it independent} of
the flavor and CP violations,  and also
of the dynamical origin of SUSY breaking,
which is a feature not shared by
the previous schemes.
This can be achieved if the symmetry which provides a rationale
for flavor and CP conserving soft terms does  {\it not}
restrict the flavor, CP and SUSY breaking mechanisms  in the
underlying theory. (A well known example of this kind is
the Peccei-Quinn symmetry \cite{pq} which leads to the CP conserving
QCD without constraining at all the CP violation in
the underlying theory.)
Unfortunately, it is very difficult to realize this scenario
within the currently popular hidden sector
supergravity (SUGRA) models \cite{nilles}.
In this letter, we wish to propose an alternative  scheme
realizing  such an attractive scenario.

The key feature of our scheme is that
a (pseudo) Goldstone supermultiplet
whose decay constant $f_A$ is far below the Planck scale
(but still far above the weak scale)
plays the role of the messenger of SUSY breaking.
Since the messenger scale $f_A$ is much smaller than the Planck scale $M_P$,
the gravitino mass is also much smaller than  the size
of soft terms
although its precise value
depends upon the details of SUSY breaking.
Clearly a global $U(1)$ symmetry spontaneously broken at
$f_A\ll M_{P}$ is a necessity for our scheme.
However this global $U(1)$  does {\it not} have to be an exact symmetry
of the theory. It may rather be an accidental classical symmetry
of the renormalizable part of the underlying supergravity model and thus
could be explicitly broken by
the Adler-Bell-Jackiw (ABJ) anomaly and also by nonrenormalizable
interactions suppressed by $1/M_{P}$.
In the following, the field variables
responsible for the spontaneous violation of $U(1)$
will be called the Goldstone
sector. It will then be assumed that the Goldstone multiplet
is the only light degree of freedom (originating from the Goldstone sector)
whose mass is far less than $f_A$.

Let us now discuss some generic features of  the SUGRA models
in which a Goldstone multiplet corresponds to the messenger of
SUSY breaking.
The K\"{a}hler potential of the model can take a rather generic form.
However the superpotential $P$ is required to
have the following form:
\be
P=P_{(V,G)}+P_{SB}+...,
\ee
where $P_{(V,G)}$ denotes the {\it renormalizable}
superpotential of the visible (V) supersymmetric standard
model sector and the Goldstone (G) sector,
while $P_{SB}$ is for
the sector which provides a dynamical seed
for supersymmetry breaking (SB) {\it at scales below} $f_A$.
Here the ellipsis stands for generic nonrenormalizable terms suppressed
by $1/M_P$.
A key feature of the above superpotential  is that the
supersymmetric standard model (SSM)  and Goldstone sectors
do {\it not} couple to the SB sector through a renormalizable term
in the superpotential.
Due to the holomorphy of the superpotential,
this feature
can  easily be an automatic consequence
of the (continuous and/or discrete) symmetries of the model.
We further require that the Goldstone sector shares
gauge interactions with both the SSM and SB sector,
but the SB sector
does not carry any of  the SSM gauge charges.
Then besides nonrenormalizable SUGRA interactions,
the SB sector communicates with the SSM sector {\it only via}
the Goldstone sector as we have desired.

Of course the renormalizable part of
our SUGRA model  is  assumed to be invariant under
a  global $U(1)$ which is spontaneously broken at $f_A$ by the
the Goldstone sector.
Since the Goldstone sector scale $f_A$ is far above the weak scale,
$P_{(V,G)}$ must be carefully tuned in order for the SSM sector
not spoiled by the large value of $f_A$.
This may be achieved by a fine tuning of some parameters
in $P_{(V,G)}$, which is technically natural
in supersymmetric model.
A more  attractive possibility would be that
$P_{(V,G)}=P_V+P_G$, i.e. the visible SSM sector and
the Goldstone sector do not have any renormalizable
superpotential coupling again
as a consequence of
the symmetries of the model.

At scales below $f_A$, the light degrees of freedom
of the model  would contain
the Goldstone chiral superfield $A$ (from
the Goldstone sector) and the SSM sector fields together with
some  SUSY breaking fields originating
from the SB sector.
Below   $f_A\ll M_{P}$,
one can safely
ignore supergravity interactions suppressed by $1/M_{P}$ and
describe physics by a globally supersymmetric effective lagrangian.
With the properties of the model discussed above,
the effective lagrangian  can be
divided into two pieces:
\be
{\cal L}_{\rm eff}={\cal L}_{\rm SB}+{\cal L}_{{\rm SSM}},
\ee
where ${\cal L}_{SB}$  is the part of the lagrangian
describing SUSY breaking dynamics which is {\it independent} of  the SSM
sector,
and ${\cal L}_{\rm SSM}$ is the part depending
on the SSM fields and the Goldstone multiplet.

Due to the (approximate) $U(1)$ invariance of the underlying
SUGRA model, the effective lagrangian of eq. (2) is constrained by the
global $U(1)$ symmetry nonlinearly realized
as:
\be
A\rightarrow A+i\alpha,  \quad \theta\rightarrow e^{ik\alpha/2}\theta,
\ee
where $\alpha$ and $k$ are real parameters
and $\theta$ denotes the Grassman coordinate of the superspace.
Here the Goldstone superfield $A$ is defined as being {\it
dimensionless}.
Note that we can always define the nonlinear symmetry
in a way that all superfields other than the Goldstone multiplet
are invariant. In this prescription,
the matter multiplets in the effective theory
would be  related to those in the underlying SUGRA model
by an appropriate
$A$-dependent field redefinition.
Note also that  $k\neq 0$ iff the above $U(1)$ is a R-symmetry.

So far, we have discussed some generic
features of the models  in which a Goldstone multiplet
plays the role of the messenger of SUSY breaking.
An interesting feature of this kind of models
is that  the shape of soft terms
in the SSM sector
is rather {\it independent} of the further
details of the model, particularly of the detailed mechanism
of SUSY breaking which would be described by ${\cal L}_{SB}$.
This is nice since the SUSY breaking dynamics
is in fact the most model-dependent but the least-understood part
of generic supersymmetric theories.

Since the shape of soft terms is of our primary interest,
in this paper we concentrate on ${\cal L}_{SSM}$ with the assumption that
the auxiliary component
$F_A$ of
the Goldstone multiplet develops an appropriate vacuum value
of order $10$ TeV through its couplings in
${\cal L}_{SB}$.
Generically  ${\cal L}_{\rm SSM}$ can be written as
\begin{eqnarray}
{\cal L}_{\rm SSM}&&=
\int d^2\theta \, [\frac{1}{4}f_a(A)W_aW_a+
P_{\rm eff}(A,\Phi_i)]
\nonumber \\
&&+
\int d^2\theta d^2\bar{\theta} \, \frac{1}{2}
Z_{i\bar{j}}(A,\bar{A})\Phi_i\bar{\Phi}_{\bar{j}}+{\rm h.c},
\end{eqnarray}
where $\Phi_i$ and $W_a$ denote the chiral matter and gauge superfields
in the SSM sector, including the quarks, leptons, Higgs doublets,
and the $SU(3)_c\times SU(2)_L\times U(1)_Y$ gauge multiplets.
Then the (anomalous) nonlinear symmetry of eq. (3) implies
\be
Z_{i\bar{j}}=Z_{i\bar{j}}(A+\bar{A}),
\quad
f_a=\kappa_a+\frac{1}{4\pi^2}c_a A,
\ee
where $\kappa_a$'s are the complex constants
whose real parts correspond to the tree level gauge coupling constants,
while $c_a$'s are real constants determined by the one-loop ABJ anomalies
for the $U(1)$ symmetry.
The effective superpotential is constrained also as
\begin{eqnarray}
P_{\rm eff}&=&e^{k^{\prime}A}\tilde{\mu}H_1H_2+e^{kA}
\tilde{\lambda}_{ijk}
\Phi_i\Phi_j\Phi_k
\nonumber \\
&\equiv&
\mu H_1H_2+\lambda_{ijk}\Phi_i\Phi_j\Phi_k,
\end{eqnarray}
where $H_1$ and $H_2$ denote the two Higgs doublets
in the SSM, and  $\tilde{\mu}$ and $\tilde{\lambda}_{ijk}$
are {\it $A$-independent} constants.

The above results on $Z_{i\bar{j}}$, $f_a$, and the Yukawa terms
are simply due to the invariance under
the nonlinear $U(1)$ transformation (3), which
is the low energy manifestation of a spontaneously broken
classical $U(1)$
symmetry of the {\it renormalizable} part of the underlying
SUGRA model.
However about the $\mu$-term, one needs a further consideration
since it can arise from the {\it nonrenormalizable}
part.
Let $Q_i$ denote the $U(1)$ charge of
the SSM multiplet $\tilde{\Phi}_i$
in the underlying
SUGRA model in which the $U(1)$ symmetry is linearly realized
as:
\be
\tilde{\Phi}_i\rightarrow e^{iQ_i\alpha}\tilde{\Phi}_i.
\ee
Clearly if $Q_{H_1H_2}\equiv Q_{H_1}+Q_{H_2}$
differs from $k$,
the SUGRA superpotential of eq. (1)
does not contain a bare $\mu$-term.
Then our $U(1)$ symmetry
can be useful for explaining
why $\mu$
is not given by  the Planck scale.
If $Q_{H_1H_2}=0$, the K\"{a}hler
potential can contain a term like  $H_1H_2$ leading to a $\mu$-term
in $P_{\rm eff}$ \cite{giudice}.
However the  size
of the resulting $\mu$
is of order the gravitino mass which is much smaller than the weak
scale in our case.
Still  the $\mu$-term in $P_{\rm eff}$
can be induced by the following form of
{\it nonrenormalizable} terms in
the SUGRA superpotential \cite{kim}:
\be
\Gamma_{(\mu)}=X_{(\mu)}H_1H_2/M^n_P,
\ee
where $X_{(\mu)}$ is a $d=n+1$ operator.
If a nonzero vacuum value of $X_{(\mu)}$ is developed
by the Goldstone sector dynamics,
it is expected that $\mu$ is of order $f_A^{n+1}/M^n_P$, and
thus of order the weak scale for $f_A=O(10^{(18n+2/n+1)})$
GeV.
Depending upon the model, not a single but
several independent $\Gamma_{(\mu)}$'s can
give comparable contributions to the $\mu$-terms. In such  case,
we {\it assume} that
those $\Gamma_{(\mu)}$'s
have a  common $U(1)$ charge
$k^{\prime}$.
We stress here that this is {\it not} a fine tuning of continuous
parameters, but a rather mild condition since those $\Gamma_{(\mu)}$'s
would already be severely constrained
by the continuous and discrete gauge symmetries of the model.
Although we have discussed the possible
role of the $U(1)$ symmetry and also of the Goldstone sector
for the $\mu$-problem,
we do not require here that it is
necessarily the case. For instance, the bare $\mu$-term
in the SUGRA superpotential
can be forbidden not by our $U(1)$ but by other means,
and thus it is possible that $Q_{H_1H_2}=k$.
It is also possible that $\langle X_{(\mu)}\rangle$ is
not developed by the Goldstone sector dynamics, but by other dynamics,
allowing $f_A$ to take a value in a rather wide range.

It is now straightforward \cite{dilaton} to find the following soft
parameters from ${\cal L}_{SSM}$ of eq. (4):
\begin{eqnarray}
m^2_{i\bar{j}}&=&|F_A|^2(Z^{k\bar{l}}\partial_AZ_{i\bar{l}}\partial_{\bar{A}}
Z_{k\bar{j}}-\partial_A\partial_{\bar{A}}Z_{i\bar{j}}), \nonumber \\
A_{ijk}&=&e^{-kA}F_A
(k\tilde{\lambda}_{ijk}-Z^{l\bar{m}}\partial_AZ_{\bar{m}(i}
\tilde{\lambda}_{jk)l}),
 \nonumber \\
B&=&F_A(k^{\prime}-\partial_A\ln(Z_{H_1\bar{H}_1}Z_{H_2\bar{H}_2})),
\nonumber \\
m_a&=&F_A\frac{\alpha_a}{2\pi}
(c_a-\sum {\rm tr}(T_a)^2
\partial_A\ln {\rm det}[Z]),
\end{eqnarray}
where $m^2_{i\bar{j}}$, $A_{ijk}$, and  $B\mu$
denote the soft scalar mass, trilinear coefficient, and bilinear
Higgs mixing coefficient respectively, while
$m_a$ is the gaugino mass.
Here the expression of the gaugino mass is valid only up
to one loop approximation, which is enough for our purpose.
Note that the above soft parameters are the
running parameters whose renormalization point is set up
by that of the renormalized
$Z_{i\bar{j}}$
and $\alpha_a$.

Using the above results,
one can  compute soft parameters
once
the $A$-dependence of
$Z_{i\bar{j}}$ is known.
In fact, without knowing much about $Z_{i\bar{j}}$,
we can extract some important properties of soft terms.
For instance, for generic $Z_{i\bar{j}}=Z_{i\bar{j}}(A+\bar{A})$,
we find
${\rm Arg}(m_a)={\rm Arg}(B)$.
Also if $Z_{i\bar{j}}=\delta_{i\bar{j}}Z_i(A+\bar{A})$, we have
${\rm Arg}(m_a)={\rm Arg}(A_{ijk}/\lambda_{ijk})$.
These two properties  guarantee
that soft terms do not give any dangerous
CP violation, e.g. a too large electric dipole moment of the
neutron,
even when the superparticle
masses are of order 100 GeV \cite{dugan}.

In the above, we could conclude that
soft terms are CP conserving
without knowing much about
$Z_{i\bar{j}}$.
In fact,
we can go far further by computing
$Z_{i\bar{j}}$
in perturbation theory.
To proceed, let us first note that
the K\"{a}hler potential of the underlying SUGRA model can always be
expanded as $K=\tilde{\Phi}^{\dagger}_i\tilde{\Phi}_i+...,$
where $\tilde{\Phi}_i$'s  denote the SSM multiplets which transform
as (7) under the $U(1)$.
Then at  {\it tree approximation}, comparing this
K\"{a}hler potential and the linear transformation (7) with the effective
lagrangian (4) and the nonlinear transformation (3) respectively,
one easily finds $\tilde{\Phi}_i=e^{Q_iA}\Phi$, and thus
\be
Z^{({\rm tree})}_{i\bar{j}}=e^{Q_i(A+\bar{A})}\delta_{i\bar{j}}.
\ee
Applying this for (9), one also finds  that all soft parameters do
vanish at  tree approximation except for $B=(k^{\prime}-Q_{H_1H_2})F_A$.
(Here we have used  the $U(1)$-invariance of
the Yukawa couplings, viz $k=Q_i+Q_j+Q_k$ for $\lambda_{ijk}\neq 0$.
Note that the formula for the gaugino masses show explicitly
that they are essentially loop effects.)

Let us now consider loop corrections to $Z_{i\bar{j}}$.
In order to see one-loop corrections,
let us write down the classical lagrangian
in terms of the multiplets $\tilde{\Phi}_i$ with
the canonical kinetic terms.
Obviously then all tree level gauge and Yukawa couplings are
$A$-{\it independent}, and thus there is
no one-loop
modification to the $A$-dependence of $Z_{i\bar{j}}$.
As a result, $m^2_{i\bar{j}}$ and $A_{ijk}$ are still vanishing,
but nonzero $m_a$'s are  given by eq. (9) with $Z_{i\bar{j}}=
Z^{({\rm tree})}_{i\bar{j}}$.
One can now include two loop corrections to
$Z_{i\bar{j}}$, but in regard to the associated
soft parameters, the results correspond mainly to the
renormalization group running of following
boundary values given at
{\it  the renormalization point} $f_A$:
\begin{eqnarray}
&&m^2_{ij}=A_{ijk}=0,
\quad m_a=
\frac{\alpha_a}{2\pi} \tilde{c}_aF_A, \nonumber \\
&&B=(k^{\prime}-Q_{H_1H_2})F_A,
\end{eqnarray}
where $\tilde{c_a}=(c_a-\sum Q_i{\rm tr}(T_a)^2)$
is determined by the ABJ anomaly due to the Goldstone sector
and $k^{\prime}$ is the $U(1)$ charge of
the operator $\Gamma_{(\mu)}$ inducing the $\mu$-term.

The above results on $m^2_{i\bar{j}}$, $A_{ijk}$ and $m_a$
are  essentially  the predictions of generic models
in which a Goldstone multiplet is the messenger of SUSY breaking.
Although   $B$ is not determined by the symmetry
consideration alone, but
depends upon the origin of the $\mu$-term,
it can also be fixed  (at least for the case that
the SSM sector is the minimal one) by the following reasoning.
Let us first note that
$Q_{X}\equiv(k^{\prime}-Q_{H_1H_2})$ corresponds to the $U(1)$ charge
of the operator ${X}_{(\mu)}$ in eq. (8) whose vacuum value
generates the $\mu$-term.
As a result, any {\it nonzero} value of $Q_{X}$
would be of order unity. The resulting
$B$ which is of order $F_A$
is then too large to have the radiative electroweak
breaking with a stable vacuum  in the context of the minimal
SSM. Note that for the boundary values of eq. (11) at $f_A$,
all soft parameters other than $B$ are of the order of $\frac{\alpha}{\pi}F_A$
at the weak scale.
This means that, at least for the case that the SSM sector
is the minimal one, ${X}_{(\mu)}$ must be
$U(1)$ invariant, i.e. $Q_{X}=0$,
leading to $B=0$
at the renormalization point $f_A$.


Interestingly enough, our
predictions  of eq. (11) are quite similar
to those of no scale SUGRA models \cite{noscale}.
They also have a common feature with the predictions
of the dynamical
SUSY breaking models of ref. \cite{dine} in the sense
that $m^2_{ij}$ at the weak scale is essentially
two loop effects mediated by the one loop gaugino masses.
However there is  a significant difference.
The messenger scale of our scheme,
i.e. $f_A$  at which our predictions are given as eq. (11),
is far below $M_P$ which is the messenger scale of no scale models,
but still far above the messenger scale $\Lambda\sim 10$ TeV
of the models discussed in \cite{dine}.
(For instance, $f_A=O(10^{10})$ GeV if it is also
the scale for the $\mu$-term generated by $d=5$ operators.)
This leads to  a sizable difference
in  the soft parameter values at the weak scale, which would
distinguish our model from the others.
More detailed low energy phenomenologies
of our model will be discussed in the subsequent work \cite{choi1}.

A nice feature of
our scheme
is that it can be implemented rather easily in a wide
class of SUGRA models.
This is mainly because the scheme
does {\it not} restrict
the flavor and CP
violations in the underlying theory and also is {\it independent}
of SUSY breaking dynamics.
Is it possible
that our scheme is realized within the low energy
limit of string theories?
Although there is a strong limitation on an exact continuous global
symmetry in string theory \cite{global}, it is still quite possible to have
an accidental global $U(1)$ symmetry whose spontaneous breaking
gives rise to the messenger Goldstone multiplet.
Another interesting  possibility
arises in models with an anomalous $U(1)$ gauge symmetry \cite{dine1}.
In such models, together with other gauge symmetries,
the nonlinear global $U(1)$ of the model independent axion
plus the anomalous gauge $U(1)$ can render an anomalous  global $U(1)$
which survives down to a scale far below $M_P$ \cite{kim1}.
One may then identify this global $U(1)$ as that
giving  the messenger Goldstone multiplet.

To conclude, we have proposed a scheme for
flavor and CP conserving soft terms in which a
Goldstone multiplet plays the role
of the messenger of SUSY breaking.
This scheme can be realized in a wide class of supergravity models
including those from string theories since
it can be implemented
{\it independently} of the nature of
SUSY breaking dynamics and also of the flavor and
CP violations in the model.
It leads to a highly predictive form of soft terms,
particularly the soft scalar masses and trilinear $A$ couplings
vanish at the scale of the Goldstone decay constant
$f_A$
whose value is around the intermediate scale.
The soft Higgs scalar mixing $B$ vanishes also at
$f_A$ at least for the case of the minimal supersymmetric
standard model.
These predictions are similar to those of no scale models,
but still leads to a distinguishable low energy phenomenology
since $f_A$ is far below the Planck scale.

\acknowledgments
We thank G. Ross for useful discussions.
This work is supported in part by KOSEF,
and the Basic Science Research Institute Program, Ministry of
Education, BSRI-95-2434.

\end{document}